\documentclass[twocolumn]{aastex62}

\usepackage{hyperref}
\usepackage{graphicx}

\hypersetup{
    colorlinks=true,
    linkcolor=blue,
    filecolor=magenta,      
    urlcolor=cyan,
}
 
\graphicspath{{./}{figures/}}

\accepted{November 15, 2022}
\submitjournal{PASP}

\usepackage{booktabs}
\usepackage{amsmath}
\shorttitle{SImMER}
\newcommand\addressresponse[1]{{#1}}
\shortauthors{Savel et al.}
\begin{document}

\title{SImMER: A Pipeline for Reducing and Analyzing Images of Stars}

\correspondingauthor{Arjun B. Savel}
\email{asavel@umd.edu}

\author[0000-0002-2454-768X]{Arjun B. Savel}
\affiliation{Flatiron Institute, Simons Foundation, New York, NY, USA}
\affiliation{Department of Astronomy, University of Maryland, College Park, MD, USA}
\affiliation{Department of Astronomy, University of California, Berkeley, Berkeley, CA, USA}

\author[0000-0001-8058-7443]{Lea A. Hirsch}
\affiliation{Kavli Institute for Particle Astrophysics and Cosmology, Stanford University, Stanford, CA, USA}
\affiliation{Department of Chemical and Physical Sciences, University of Toronto Mississauga, ON, CA}

\author[0000-0001-6171-7951]{Holden Gill}
\affiliation{Department of Astronomy, University of California, Berkeley, Berkeley, CA, USA}

\author[0000-0001-8189-0233]{Courtney D. Dressing}
\affiliation{Department of Astronomy, University of California, Berkeley, Berkeley, CA, USA}

\author[0000-0002-5741-3047]{David R. Ciardi}
\affiliation{NASA Exoplanet Science Institute-Caltech/IPAC Pasadena, CA 91125, USA}
% \nocollaboration{2}



\begin{abstract}
We present the first public version of \texttt{SImMER}, an open-source \texttt{Python} reduction pipeline for astronomical images of point sources. Current capabilities include dark-subtraction, flat-fielding, sky-subtraction, image registration, FWHM measurement, contrast curve calculation, and table and plot generation. \texttt{SImMER} supports observations taken with the ShARCS camera on the Shane 3-m telescope and the PHARO camera on the Hale 5.1-m telescope. The modular nature of \texttt{SImMER} allows users to extend the pipeline to accommodate additional instruments with relative ease. One of the core functions of the pipeline is its image registration module, which is flexible enough to reduce saturated images and images of similar-brightness, resolved stellar binaries. Furthermore, \texttt{SImMER} can compute contrast curves for reduced images and produce publication-ready plots. The code is developed online at \url{https://github.com/arjunsavel/SImMER} 
and is both pip- and conda-installable. We develop tutorials and documentation alongside the code and host them online. With \texttt{SImMER}, we aim to provide a community resource for accurate and reliable data reduction and analysis.

\end{abstract}

\keywords{Astronomy data analysis --- 
Astronomy data reduction --- Photometry --- Observational astronomy}

%% We recommend that authors also use the natbib \citealtp
%% and \citealtt commands to identify citations.  The citations are

\section{Introduction}
The reliable conversion of raw images into science-ready data is a crucial step in any observational project. If performed incorrectly, errors can propagate throughout the remainder of the project, in some cases significantly affecting scientific conclusions.

With the goals of reproducibility and accuracy in mind, open-source data reduction pipelines offer a number of clear benefits. If the entire relevant community is able to view a piece of code, researchers can verify the work of others and identify bugs. Accordingly, the code can serve as a stable jumping-off point, allowing researchers to move toward answering their science questions more quickly with heightened trust in their underlying calculations. The continuous development and well-defined feature-addition of open-source projects can furthermore  help integrate early-career researchers into the astronomical research community, providing contained projects with concrete value in conjunction with the opportunity to rapidly interface with real data. 

Despite the benefits of open-source and accessible data reduction pipelines, the time investment required to develop this software results in it not being guaranteed for a given instrument. Even if every instrument included open-source reduction pipelines, the data reduction routines that some observatories offer are often implemented in proprietary software like IDL (\citealt{cite-key}) or in IRAF (\citealt{tody1986iraf}, \citealt{tody1993iraf}), which is no longer actively developed or maintained.\footnote{\url{https://iraf-community.github.io/}} However, with the popularization of the astronomy-oriented \texttt{Astropy} package (\citealt{price2018astropy}) and the movement of undergraduate astronomy curricula toward \texttt{Python}, undergraduates and early-career astronomers already have at their disposal a variety of \texttt{Python} tools related to research work. Therefore, open-source \texttt{Python} data-reduction pipelines are an ideal addition to the astronomy stack for early-career astronomers, being written in a computing language that has a vibrant scientific ecosystem \citep{harris2020array} and that is free, eliminating the need for an expensive license. 

In this work, we present the first public release (v1.0.0) of the \texttt{SImMER} package, an open-source image reduction pipeline for point sources.  The pipeline features multiple image registration modes, allowing users to accurately and flexibly align and center images. Different registration modes are tailored to reducing images of wide binaries and/or saturated stars. With this code, we aim to produce a \texttt{Python} community tool that is well documented, well tested, modular, and adaptable. Being written in \texttt{Python}, this package aims to have strong cross-platform functionality, working well on a Mac, Linux, or Windows machine.

\addressresponse{\texttt{SImMER} is primarily meant to perform standard data-reduction steps (dark-subtraction, flat-fielding, image-registration), and basic analysis (contrast curves, aperture photometry) of images dominated by a single point source. \texttt{SImMER} currently does not have the precision to reduce data for, e.g., precise astrometry or crowded field photometry. After a brief overview of capabilities (Section~\ref{sec:overview}), this paper first provides an overview of the code's current data-reduction (Section~\ref{sec:metadata}--Section~\ref{sec:registration modes}) and analysis (Section~\ref{sec:contrast curves}) capabilities}, followed by a discussion of planned functionality (Section~\ref{plans}) and concluding remarks (Section~\ref{sec:summary}). Active development of our project takes place on GitHub,\footnote{\url{https://github.com/arjunsavel/SImMER}}
while the latest stable version of the code (v1.0.0 as of the acceptance of this paper) is distributed on PyPI\footnote{\url{https://pypi.org/project/SImMER}}
and conda-forge.\footnote{\url{https://github.com/conda-forge/SImMER-feedstock}}\\

\section{Capabilities} \label{sec:capabilities}
\subsection{Overview}\label{sec:overview}
\begin{figure*}[t]
  \centering
  \includegraphics[scale=0.13]{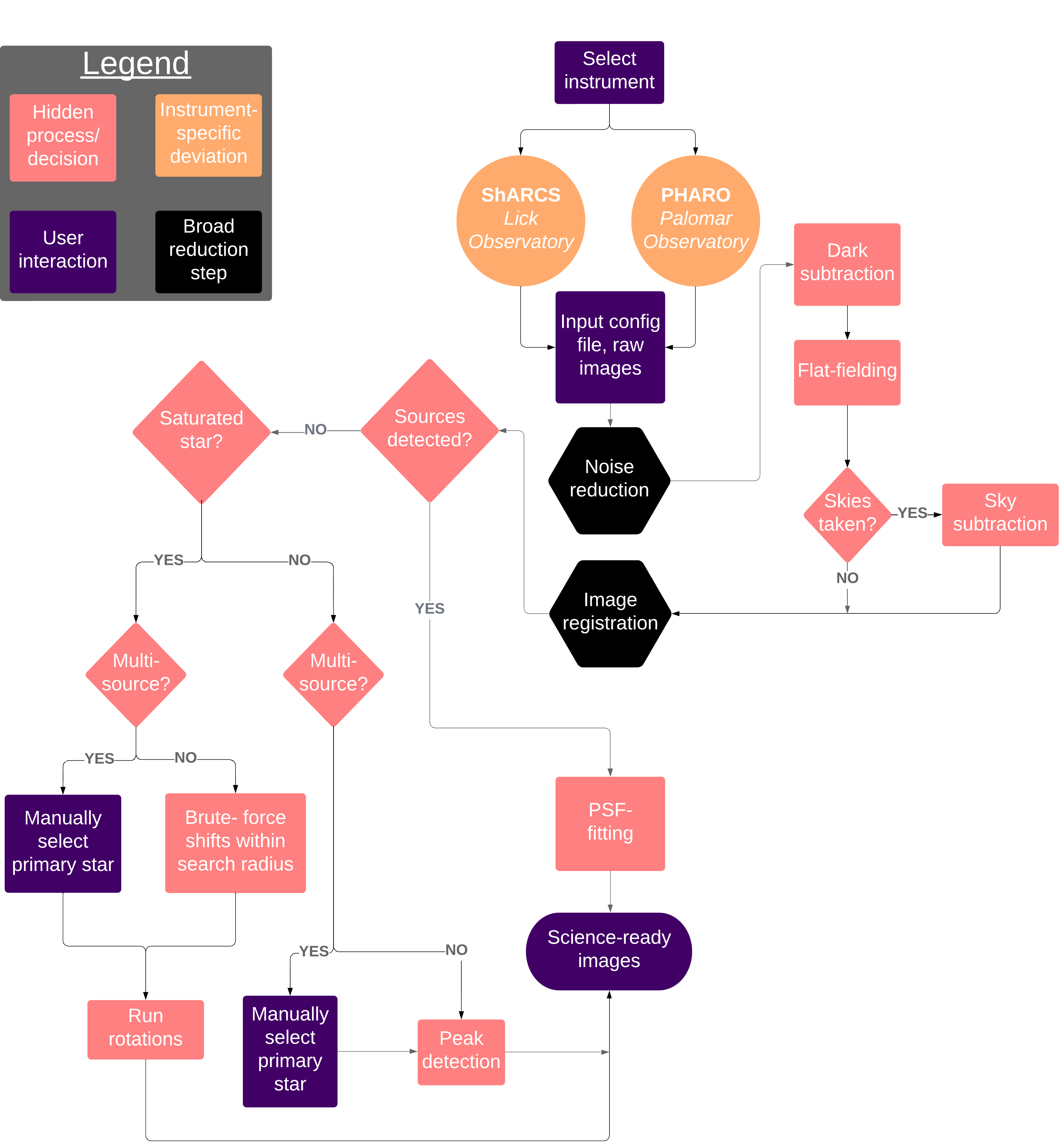}
  \caption{Flowchart for the \texttt{SImMER} reduction pipeline. Circles indicate steps that differ from facility to facility; diamonds indicate potentially diverging paths; squares indicate pipeline actions and calculations; and hexagons indicate macro steps comprised of smaller ones.}
  \label{fig:flowchart}
\end{figure*}

\texttt{SImMER} is a pipeline made to reduce photometric observations  of point sources; Figure~\ref{fig:flowchart} illustrates the steps that it takes to do so. In sum, the code performs dark subtraction to account for dark current, the signal recorded when the detector is not exposed to light; flat-fielding to account for interpixel sensitivity variations to a uniform distribution of incident photons; (optional) sky-subtraction to remove brightness on the sky unrelated to a target of interest; and aligning and combining of images that may be taken in succession to mitigate the effects of cosmic ray hits, atmospheric turbulence, and transient instrumental effects (Figure~\ref{fig:flowchart}). During the reduction process, bad pixel maps are used when possible to filter out pixels with known deviant photon responses, and intermediate plots and tables are automatically generated. After the reduction process, \texttt{SImMER} has functionality to estimate the full-width half-maxes of point sources and to compute contrast curves, which are estimates of image sensitivity (see Section~\ref{sec:contrast curves}). Each module is briefly described in the Appendix~\ref{sec:appendix}.

The pipeline's contrast curves have been benchmarked against similar codes that have been used extensively in the literature (Section~\ref{sec:contrast curves}). \texttt{SImMER}-reduced images routinely achieve contrasts of $\Delta K_S\approx6$ at 1\farcs\, and $\Delta K_S\approx8$ at 4\farcs\, on ShARCS datasets (Section~\ref{sec:sharcs}).

\subsection{\addressresponse{Required metadata}}\label{sec:metadata}
\addressresponse{\texttt{SImMER} is meant to be modular, allowing data reduction across imagers. To successfully reduce and analyze data, \texttt{SImMER} requires the following information about the instrument that gathered the data}:

\begin{itemize}
    \item \addressresponse{\textit{Instrument plate scale}, to convert contrast curves from pixel units to arcseconds 
    \item \textit{Center of illuminated imaging area}, if this is distinct from the center of the raw image (e.g., as with ShARCS\footnote{\url{https://mthamilton.ucolick.org/techdocs/instruments/sharcs/detector}})}
\end{itemize}

\addressresponse{And in the FITS header of each image:}
\begin{itemize}
    \item \addressresponse{\textit{Filter}, so that images can be matched with their corresponding flats
    \item \textit{Integration time}, so that images can be matched with their corresponding darks}
\end{itemize}

% \subsection{How to use \texttt{SImMER}} \label{howto}

\subsection{Supported instruments}
Our pipeline currently provides functionality for two instruments: the ShARCS camera on the Shane 3-m telescope at Lick Observatory \citep{kupke2012shaneao,gavel2014shaneao,srinath2014swimming,mcgurk2014commissioning} and the PHARO camera on the Hale 5.1-m telescope at Palomar Observatory (\citealt{hayward2001pharo}). The modular, object-oriented approach of our code, however, lends itself nicely to future application to other instruments (see Section~\ref{plans}).

While both currently supported instruments have adaptive optics (AO) capabilities, our code is not necessarily restricted to AO-enabled instruments. In principle, the pipeline functionality could be expanded to reduce observations of point sources from any imager, though final image quality would be dependent on the seeing and instrument performance.

We aim to provide a comprehensive set of tutorials, worked examples, and troubleshooting guidelines to the \texttt{SImMER} user community. Our documentation is built in its entirety online through Read the Docs;\footnote{\url{https://SImMER.readthedocs.io/en/latest/pages/about.html}}
  currently, the site hosts installation guides, guidance on constructing configuration files, and quickstart tutorials on reducing ShARCS (Section~\ref{sec:sharcs}) and PHARO (Section~\ref{sec:pharo}) data. All functions, methods, and classes in our code are additionally documented and automatically linked. Bugs are reported on our issues page,\footnote{\url{https://github.com/arjunsavel/SImMER/issues}} 
and recurring user troubles are recorded on our FAQ page.\footnote{\url{https://SImMER.readthedocs.io/en/latest/pages/FAQ.html}}
 As our functionality continues to grow (Section \ref{plans}), we will ensure that the available resources grow alongside it.

\subsubsection{ShARCS}\label{sec:sharcs}
The Shane Ao infraRed Camera-Spectrograph \linebreak (ShARCS; \citealt{kupke2012shaneao}, \citealt{gavel2014shaneao}) is a near-infrared camera on the Shane 3-m telescope meant to be used in conjunction with the Shane Adaptive Optics system (ShaneAO) at Lick Observatory. When reducing data from this instrument, \texttt{SImMER} uses a bad pixel map that was last updated at the commissioning of the instrument prior to installation in 2014 (Rosalie McGurk, private communication). The camera has a plate scale of 0\farcs0333 per pixel, with a field of view of 20\farcs4 x 20\farcs4. 

\begin{figure*}[t]
  \centering
  \includegraphics[scale=0.5]{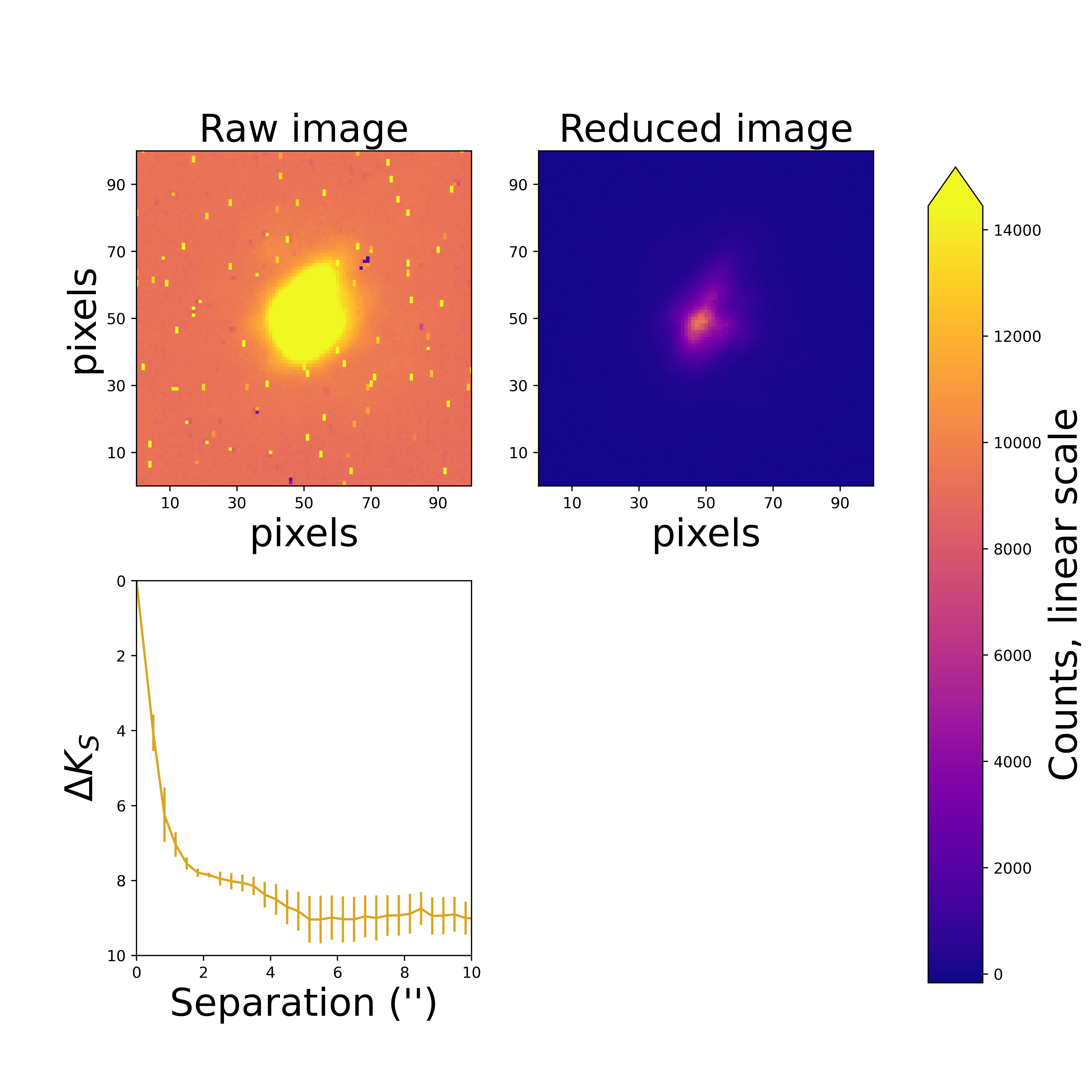}
  \caption{An example of a raw Shane image (top left) and reduced Shane image (top right), along with the final image's contrast curve (bottom left). This target is K09203794, observed on the Shane 3-m telescope on 2018-7-23. \addressresponse{These images} are \addressresponse{3}\farcs33 by \addressresponse{3}\farcs33 \addressresponse{cuts} of the sky around the target, with ICRS East pointing upward.}
  \label{fig:sharcs}
\end{figure*}

Artificial bright sources known as ``ghost images'' may be present in images produced by this instrument due to secondary reflections within the telescope, but they can be identified by their consistent position angle and angular separation across observations. An example of raw Shane data compared to reduced Shane data is depicted in Figure~\ref{fig:sharcs}. These data are images of K09203794, a star observed by the Kepler spacecraft \citep{borucki2010kepler,batalha2010selection,bryson2010kepler,haas2010kepler,jenkins2010overview,koch2010kepler} and observed on the Shane 3-m telescope on 2018-7-23. These observations were obtained on a cloud-free night with seeing of approximately 1\farcs.

\subsubsection{PHARO}\label{sec:pharo}
The Palomar High Angular Resolution Observer (PHARO) is an infrared camera \citep{hayward2001pharo} that sits behind the P3K adaptive optics system \citep{dekany2013palm} on the Hale 5.1-m telescope at Palomar Observatory. The instrument offers two plate scales: 0\farcs025 per pixel and 0\farcs040 per pixel.

There is currently not a bad pixel map specific to this instrument, so we make use of a more general NaN-filtering method that can be applied to other instruments. Benchmarking this approach against the ShARCS bad pixel map on Shane data reveals that in regions on the detector far away from a source, both bad pixel methods produce nearly identical results. However, images reduced with these two methods are slightly different in regions close to sources. These differences are generally on the order of 200 counts per pixel.

Raw PHARO images are separated into four independent quadrants, each quadrant including 512 x 512 pixels; the four parallel signal chains allow for rapid detector readout \citep{hodapp1996}. When the data are transferred to FITS files, they are written as separate extensions \citep{hayward2001pharo}. \texttt{SImMER} includes functions that read and flatten this data into single-exposure FITS files, which can subsequently be passed through the rest of the pipeline in the same manner as ShARCS data.

\subsection{Image registration modes}\label{sec:registration modes}
Accurate image registration ensures that stars are aligned and centered consistently across multiple frames, which is crucial for reaching desired signal-to-noise ratios. 

\addressresponse{For ShARCS and PHARO data, an intermediate-complexity registration solution is optimal. At the more basic end (from an image-processing perspective), using a World Coordinate System (WCS)-fitting service is not ideal when observing especially sparse fields. Our currently supported telescopes' fields of view are small enough that it is not uncommon for the target star to be the only detectable source on an image.\footnote{\addressresponse{As a test, we submitted both raw and reduced ShARCS images of the exoplanet host TOI-2015 to \url{astrometry.net} and found that the service was unable to find a solution for the target star.}} At the high-complexity end, an extremely precise registration algorithm is not warranted for our current purposes. Common science cases that use data of our quality \citep[e.g.,][]{savel2020closer} involve searching for faint stars that are close on-sky to individual stars, as opposed to, e.g., milliarcsecond-level astrometric monitoring of the Galactic Center \citep{yelda2010improving} possible with instruments such as the Hubble Space Telescope. Even unsaturated stellar targets on ShARCS can have point-spread functions (PSFs) that are misshapen and span tens of pixels due to seeing limitations (e.g., Figure~\ref{fig:sharcs}); given that we oversample the PSFs of our targets, techniques such as drizzling \citep{fruchter2002drizzle} are not required. Furthermore, errors due to optical distortions and warping from currently supported instruments and adaptive optics pipeline are likely swamped by error due to seeing; early calibration of PHARO, for instance, showed limited signs of distortion in pinhole mask images \citep{hayward2001pharo}.} 

\addressresponse{With these considerations in mind, our image registration process simply involves stacking centered images, neglecting effects such as stretches and rotations in the image.}

We currently provide three primary image registration modes tailored to differing data scenarios that necessitate discrepant treatments: ``\addressresponse{Quick-look},'' ``Saturated,'' and ``\addressresponse{multi-source}.'' Some of these primary modes can be chained together---e.g., a \addressresponse{multi-source} can be reduced with the saturated image scheme.
\subsubsection{\addressresponse{Quick-look}}
The simplest way to find the center of an image is to identify the positions of its local maxima. In the interest of utilizing a well-tested, computationally efficient, and easily integrated algorithm, we opt for iterative use of the \texttt{peak\_local\_max} function provided in the \texttt{scikit-image} library. This function first dilates a target image by applying a maximum filter (sliding a box with a user-determined size (default: 100 pixels) over the image and setting all values within the box to the maximum value within the box). In the process, nearby local image maxima are merged. The size of the box determines the separations within which maxima are considered ``nearby.'' This filtered image is then compared to the original image, and the coordinates at which the original image equals the filtered image are set as the true local maxima. A visual demonstration of this process can be found in the \texttt{scikit-image} documentation.\footnote{\url{https://scikit-image.org/docs/stable/auto_examples/segmentation/plot_peak_local_max.html}}

While the runtime of the \texttt{peak\_local\_max} function (on the order of one millisecond per image) is amenable to application to multiple images, one complication with its direct application is that a threshold must be set by the user for peak detection. This approach is not preferable when reducing data sets containing on the order of a dozen targets. In order to run this procedure automatically with minimal user intervention, we adjust the peak threshold until only two peak coordinates remain. To arrive at this condition, we implement a binary search algorithm to determine the above-mentioned threshold, reducing the search time from a worst-case time-complexity of $\mathcal{O}(n)$ to a worst-case time-complexity of $\mathcal{O}(log(n))$ for varying the threshold. In practice, this approach can result in a speed-up of more than two orders of magnitude; a single registration time decreases from 8.29~s $\pm$ 799~ms to 12~ms $\pm$ 168~$\mu$s.\footnote{Here and throughout this paper, explicit code runtimes are calculated based on runs performed on a 2017 MacBook Air (1.8 GHz Intel Core i5 processor) running macOS Mojave with 8 GB of memory.} We find that we need not place additional constraints on minimum peak distance in the peak-detection algorithm. As a caveat, we anticipate that this algorithm may not work well with \addressresponse{all} PHARO data, which may contain brightening near the edges of frames.\footnote{Future versions of \texttt{SImMER} will require that maxima are not flagged for the 10\% of pixels closest to the image edge, conservatively.} \\

\addressresponse{This method is in general approximate, and it is best suited for exploratory data reductions of large data sets. The results may be discrepant from more robust approaches (such as the ``saturated'' method; Section~\ref{sec:saturated}).}

\subsubsection{\addressresponse{Empirical PSF}}\label{sec:empirical psf}

\addressresponse{The quick-look method is very susceptible to error induced by observing conditions. On nights with particularly sub-par seeing, for instance, shot noise may strongly influence the derived center of a source. \texttt{SImMER} therefore offers a number of more image registration schemes that are more robust under noisy conditions.}

\addressresponse{The first of these image registration schemes is the ``empirical PSF'' procedure. We assume that changes in the PSF are due largely to differential inter-star (e.g., as controlled by airmass and magnitude) and inter-filter performance. Similarly, we assume that temporal variations in the PSF (e.g., due to varying cloud cover or varying AO performance) are negligible over successive exposures. With these assumptions made, we begin with the given position of the primary source on the image as input to \texttt{SImMER}. \footnote{\addressresponse{\texttt{SImMER} includes a wrapper to the \texttt{photutils} implementation of DAOFIND \citep{stetson1987find}, but users can input their own source.}} We next fit the PSF of the source in a single image using a flexible functional form modeled after a two-dimensional Gaussian:}

\begin{equation}
    F = \frac{1}{2\pi s_{x'}s_{y'}}e^{(-x'^2/2s_{x'}^2 -y'^2/2s_{y'}^2)},
\end{equation}

\addressresponse{where $F$ is the modeled flux of the source, $s_{x'}$ controls the standard deviation in the $x'$ axis, $s_{x'}$ controls the standard deviation in the $y'$ axis, and $x'$ and $y'$ are rotated by an angle $\phi$ from the $x$ and $y$ axes, respectively. $x'$ and $y'$ are also shifted from the origin by a tunable distance in each axis.}

\addressresponse{We jointly fit the PSF of the central source across successive exposures as follows. Our fitting procedure, using the \texttt{emcee} package \citep{emcee}, produces posteriors distributions for each PSF parameter for each exposure. To assess the joint probability distributions of the shared PSF parameters (i.e., $s_{x'}$, $s_{x'}$, and $\theta$) under the assumption of independent measurement, we multiply the posterior distributions of each parameter across successive exposures \citep[e.g.,][]{brogi2017framework}. True simultaneous fitting would likely require the usage of, e.g., nested sampling \citep{skilling2006nested}, which can before better than standard Markov Chain Monte Carlo algorithms for strongly multimodal and/or degenerate parameter spaces \citep[e.g.,][]{feroz2013exploring}.}

\addressresponse{With the PSFs and their centers fit, the source centers of each image are more precisely found with a \texttt{DAOFind} \citep{stetson1987find} call tuned to the fit PSF. Registration proceeds by stacking images according to their centers. This approach can cleanly accommodate rotations between successive exposures (in the $\phi$ parameter).}

\subsubsection{Saturated}\label{sec:saturated}
An observer may need to reduce saturated data for a variety of reasons. Some science goals necessitate bright star observations---e.g., a search for very faint stellar companions that are fairly distant (e.g., 5\farcs to 10\farcs) from a bright (e.g., $R$ = 7 mag) primary in a broadband filter (e.g., $J$ band). For instance, the science case in \cite{hirsch2021understanding} required such a saturation of bright primaries in order to maximize the observed brightness contrasts. In other cases, an exposure might have inadvertently saturated due to an incorrectly estimated exposure time or variable cloud cover.

For saturated targets, the aforementioned peak-finding algorithm may not operate as intended, even for Shane data lacking fringe brightening. In particular, the saturated region may be large enough that multiple spots in a star's point-spread function (PSF) would match with the dilated image, were the dilation size smaller than the stellar PSF. Even if the dilation size were larger than the stellar PSF, the center of the saturated star's PSF would not necessarily be chosen as the ``peak.'' 

\begin{figure}[t]
  \centering
  \includegraphics[scale=0.66]{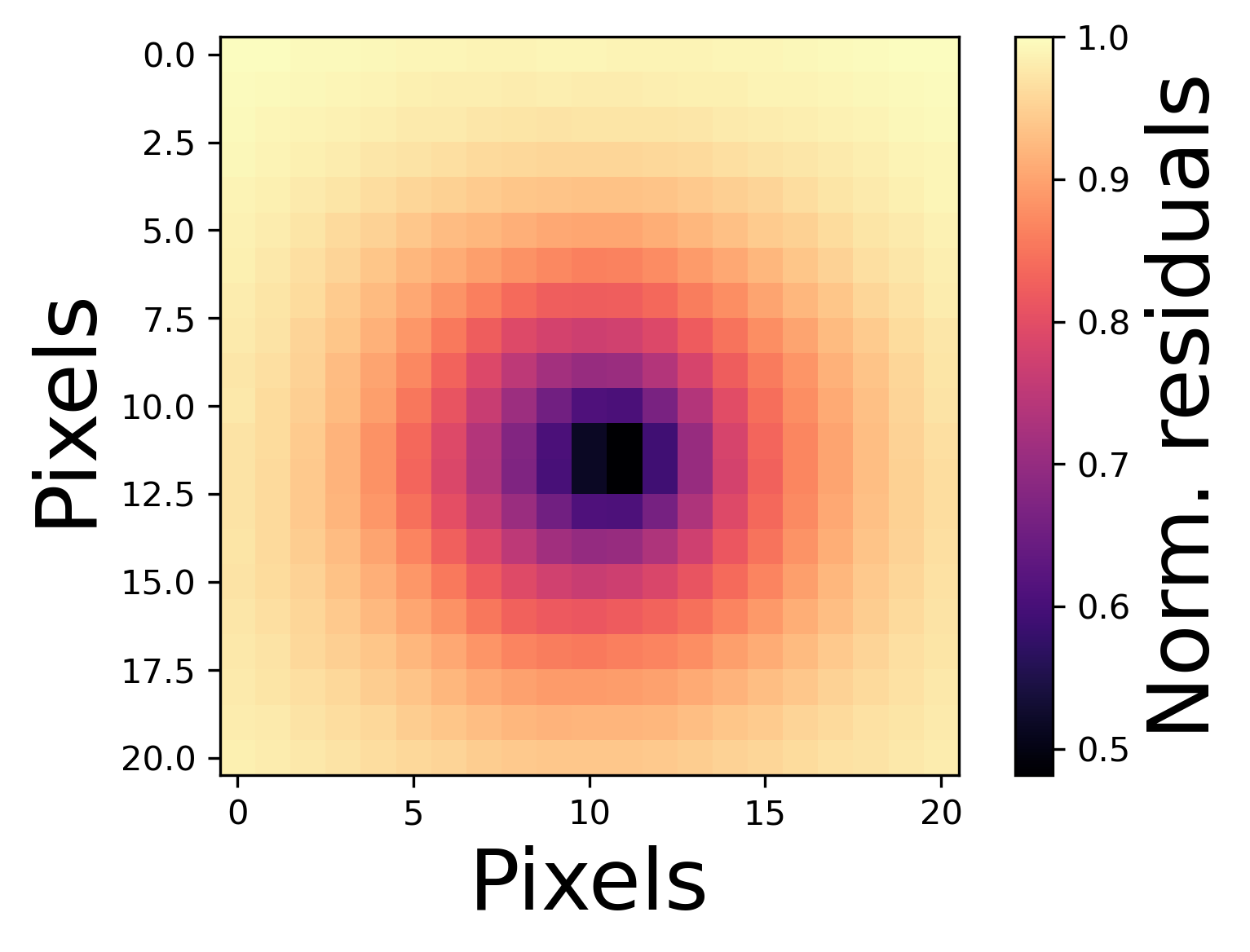}
  \caption{A visualization of the rotations step performed for the star K09203794. For each pixel within the search box, the image is rotated by $90 \degr$ three times in succession. Each rotated image is subtracted from the original image, and the results from all three rotations are added together to arrive at the total residual value for that pixel (the colorbar for this image). Residuals are normalized on the interval [0, 1].}
  \label{fig:rots}
\end{figure}

With these constraints in mind, we adapt a registration procedure with sub-pixel accuracy detailed in \cite{morzinski2015magellan} for saturated stars. First, we choose a naive center of the star: the center of the raw image. We then incrementally rotate the image around this center, with rotation angles $\theta$ spanning $0^{\circ}$ to $360^{\circ}$. Finally, we subtract these rotated images from the original image and sum the residuals from all subtractions. We subsequently change the image's center of rotation prior to repeating the process, as in Figure~\ref{fig:rots}. We compute these residual sums over a search box of centers of rotation, taking the center of rotation corresponding to the lowest summed residuals to be the center of our star. 

Mathematically, our approach minimizes the residual function

\begin{equation}
    R(x_{\text{center}}, y_{\text{center}}) = \sum\limits_{\theta=0}^{360} (I_{\text{rot}}(\theta) - I_0),
\end{equation}

varying $(x_{\text{center}}, \,y_{\text{center}})$ over a search box.

\begin{figure}
    \centering
    \includegraphics[scale=0.25]{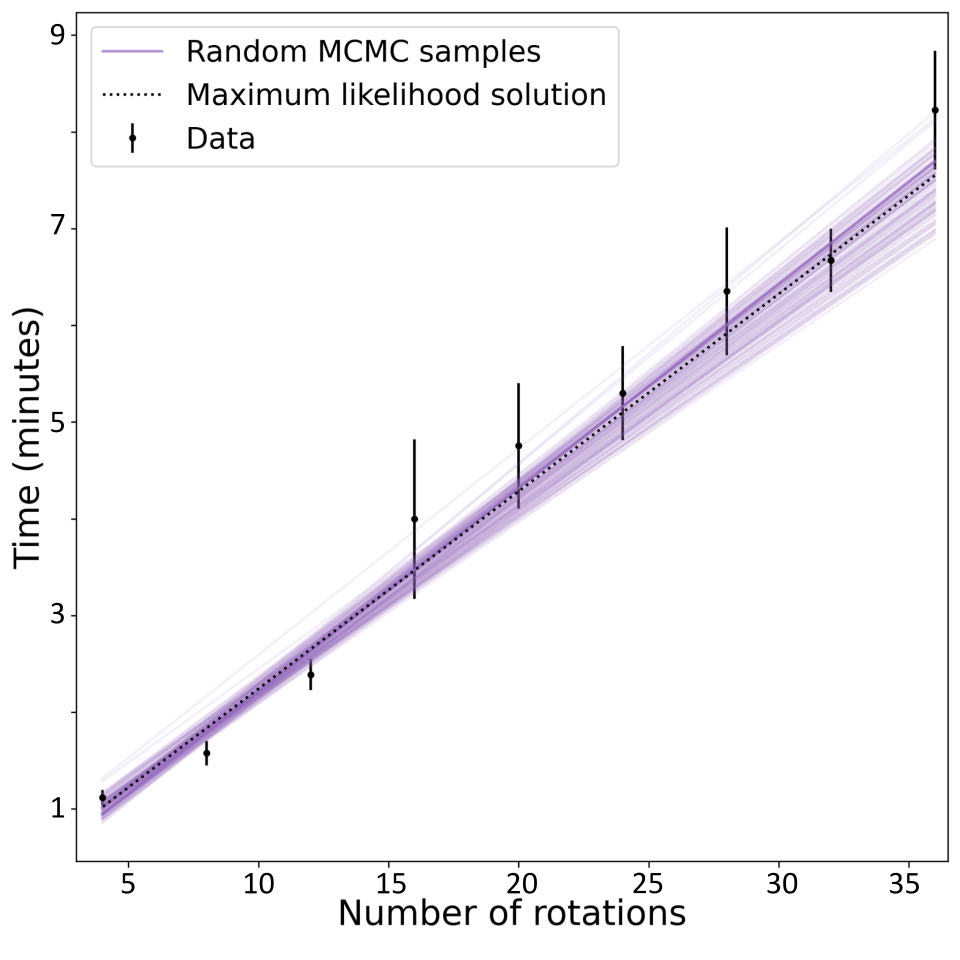}
    \caption{The runtime of the \texttt{SImMER} saturated image registration mode as a function of the number of rotations performed within the algorithm. Note that the intercept of this slope is system-dependent, but the runtime growth is system-independent.}
    \label{fig:runtime}
\end{figure}

In the original implementation, \cite{morzinski2015magellan} rotated the image in $10^{\circ}$ increments from $5^{\circ}$ to $355^{\circ}$. Reducing the number of rotations to three significantly decreases the runtime of our algorithm; as demonstrated in Figure~\ref{fig:runtime}, the runtime increases linearly with the number of rotations, so we increase the speed of our program by a factor of ~7. In the interest of symmetry, the rotation angles that we choose are $90^{\circ}$, $180^{\circ}$, and $270^{\circ}$. Additionally, the choice of these rotation angles avoids the need for any interpolation of image data; this approach is advantageous because it both reduces the number of computations performed and removes a source of potential numerical error. 

To verify the growth of our runtime, we first run the image registration function 10 times at various numbers of rotations; the standard deviation of these timings is taken to be the error in runtime at that number of rotations. These data are then fit by minimizing the negative log likelihood of our function. Next, we use the \texttt{emcee} (\citealt{emcee}) package to sample the posterior distribution about this likelihood maximum. Using the standard stretch sampler move, we run 32 independent chains over 50,000 steps, achieving convergence by checking the integrated autocorrelation time of the chain (\citealt{goodman2010ensemble}) and the Gelman-Rubin statistic (\citealt{gelman1992inference}). We then repeat this procedure for a quadratic fit. We find that the Bayes factor favors a linear model of runtime growth to a quadratic one 24:1---which, by the criteria of \cite{kass1995bayes}, is ``strong'' evidence against the quadratic model.

Of the image registration modes provided (\addressresponse{quick-look}, saturated, and multi-source), we determine the saturated mode to be the most robust, though there is a trade-off with speed; this method can take on the order of 2.5 minutes to execute for a single star, as opposed to the roughly 10~ms required for the \addressresponse{quick-look} method. We will further develop \texttt{SImMER} to reduce the runtime of this method (see Section \ref{plans}).
\subsubsection{\addressresponse{Multi-source}}
If a user is observing wide stellar binaries of similar apparent magnitudes, measured brightness fluctuations unrelated to astrophysical phenomena (in particular, atmospheric turbulence) may cause one star to appear brighter than the other in one frame and fainter than the other in a successive frame. Left untreated, this effect would result in the previously described algorithms attempting to center successive frames of the same target around different stars, disrupting the image registration process. This problem would be exacerbated in observing strategies that make use of wide dither boxes or if the stars are separated by more than a substantial fraction of the image. For instance, if the target star were centered in the image and the companion near the image's edge, a naive reduction could center some images on the companion star and produce strong edge artifacts in the reduced image.

To rectify this issue, we include in our pipeline a \addressresponse{multi-source} mode. In this mode, the user selects in each image the star that they would like to designate as the primary. After the user clicks on the rough photocenter of their desired primary, the pipeline performs a search restricted to that region. With this being a decidedly more hands-on approach to image reduction, it should be noted that this mode is only recommended to be used when necessary.

\subsection{Contrast Curves}\label{sec:contrast curves}
\begin{figure}
    \centering
    \includegraphics[scale=0.46]{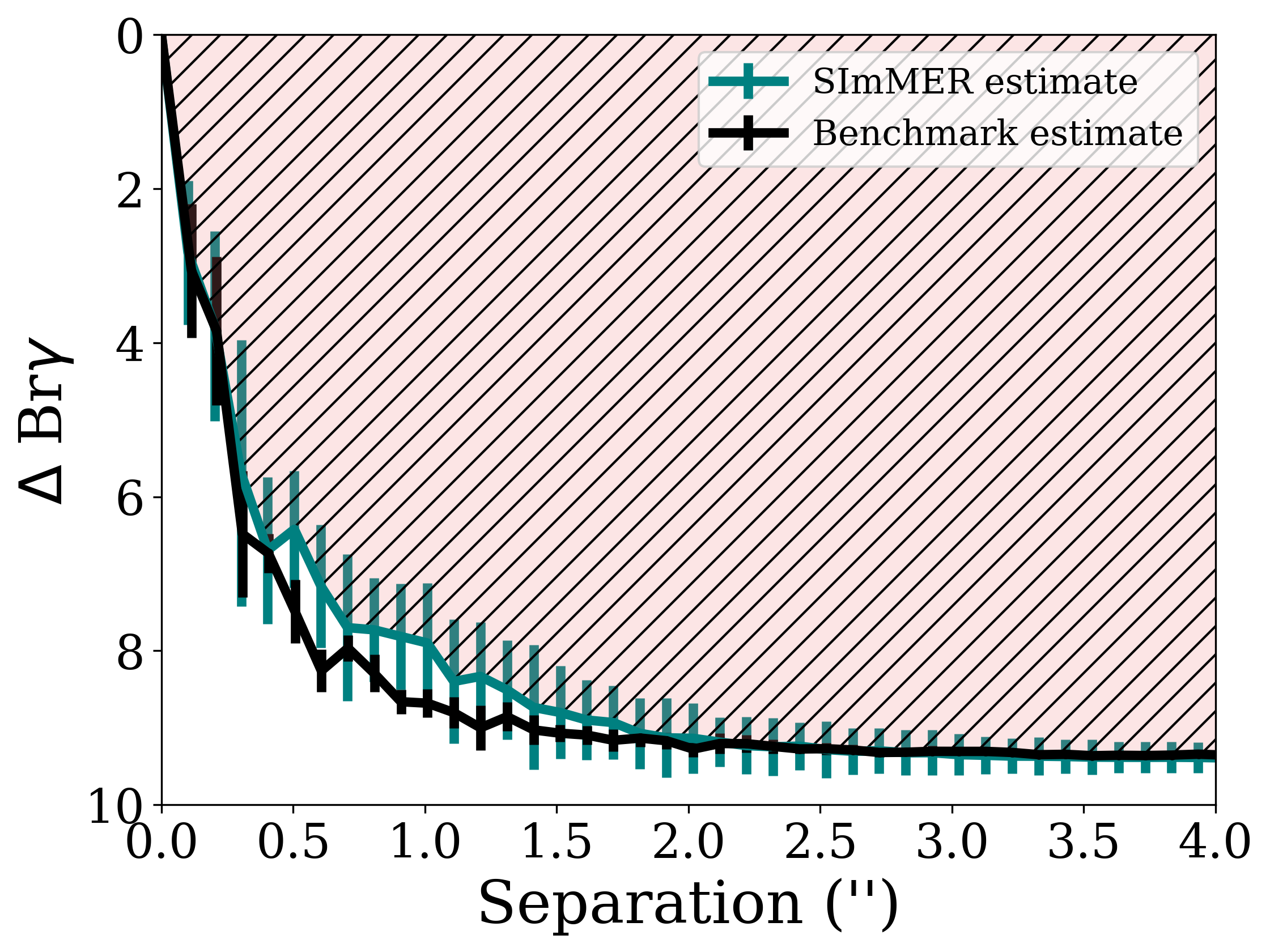}
    \caption{Example contrast curve, computed radially from TOI 1774 with a FWHM of 3.1 pixels. The benchmark estimate is calculated with the code used in, e.g., \cite{ciardi2017k2}; its estimates are consistent with the \texttt{SImMER} estimate within 1$\sigma$. For illustrative purposes, the region of parameter space in which the data are sensitive to stellar sources is hatched. This data image data and benchmark contrast curve results are available on ExoFOP.}
    \label{fig:contrast_plot}
\end{figure}

After using \texttt{SImMER} to produce final, reduced images, users can apply the code's contrast curve module to estimate the sensitivity of a search for sources on an image (Figure~\ref{fig:contrast_plot}). Contrast curves calculate the detection limit for stellar sources in concentric annuli centered on a target star, reducing two spatial axes to a single one (radial separation from the target star). In these calculations, contrast is measured relative to the brightness of the target star. This module is based on a contrast curve code that has been used in a number of searches for nearby stellar companions to exoplanet host stars over the past decade \citep[e.g.,][]{ciardi2017k2}.

The steps our code takes to compute contrast curves are as follows:

\begin{enumerate}
    \item Estimate the FWHM of the primary star in the image.
    \item Perform aperture photometry (with the \texttt{photutils} package; \citealt{larry_bradley_2020_4044744}) on the image within a FWHM of the image's center to determine the total photon counts from the target star ($A_{\rm central}$). 
    \item Construct concentric annuli around the target star, which is centered on the image. Each annulus has a width equal to the FWHM of the target star. 
    \item Split each annulus $i$ into $j$ wedges. The wedges are of equal angular width and are used to estimate dispersion in the aziumthually dependent sensitivity for a given radial distance. The exact number of wedges is not crucial for maintaining the algorithm's accuracy; we find that $j=12$ samples each annulus reasonably well.
    \item Calculate the mean and standard deviation of the photon counts for the pixels in each wedge of each annulus.
    \item For each annulus of each wedge, use the mean and standard deviation to simulate a noisy image. Insert a Gaussian source with a FWHM equal to the FWHM of the primary star and a signal-to-noise ratio of 5 into the simulated image.
    \item Perform aperture photometry on the simulated image to determine the photon counts for the simulated source ($A_{ij}$).
    \item For each annulus $i$, take the contrast value at the separation corresponding to $i$ to be 
    \begin{equation}
        \frac{1}{j}\sum^{j}A_{ij}.
    \end{equation}
   Take the standard deviation of all $A_{ij}$ as the error of the calculation at the separation corresponding to $i$.
    \item Convert to log space and subtract $A_{\rm central}$ to find the contrast in magnitudes at each separation. 
\end{enumerate}

By sampling each annulus $j$ times, this approach accounts for potential azimuthal variations of detection sensitivity in the reduction of information to a single dimension (radial separation from the primary star). 

Potential companions can be identified in our contrast curves as regions of higher contrast error, as the companion affects only a subset of the simulated images, raising the standard deviation across wedges. To increase contrast curve accuracy, users can manually specify centers of the annuli and exclude potentially anomalous pixel values (e.g., pixels where the value is higher than 10 standard deviations above of its annulus, while not belonging to a potential star).

The contrast curve results from \texttt{SImMER} are in good agreement with the previously published results from \cite{ciardi2017k2}.  We do note some $\lesssim 1\sigma$ differences between the two algorithms, which may be result from differences in the sub-pixel registration algorithms in original IDL code \citep{ciardi2017k2} and the new \texttt{Python}-based \texttt{SImMER} presented here.

\section{Planned functionality} \label{plans}
The framework that we have constructed in this first public version of \texttt{SImMER} is modular and extendable. With the pipeline being actively developed, we plan in the near future to implement the below features, which are tracked on our issues page\footnote{\url{https://github.com/arjunsavel/SImMER/issues}} 
and tied to specific releases on our projects page.\footnote{\url{https://github.com/arjunsavel/SImMER/projects}}

\renewcommand{\labelitemi}{---}
\begin{itemize}
    \item \textbf{Tracking WCS coordinates when stacking images.}
    The FITS headers corresponding to raw data from ShARCS and PHARO contain the right ascension (RA) and declination (Dec) of the image center. The header of the FITS output of this code does not currently track how the image has been shifted; in the future, the code will track image shifts with respect to WCS and write shifted RA and Dec coordinates to the final FITS file.
    \item \textbf{Creating a reduction mode that works directly from raw data without a config file.} The information required for the config file is often contained in FITS headers. If our users are sure of the FITS files headers, they will be able to call this reduction mode on any directory containing raw data. Initial work toward this goal has already commenced, as \texttt{SImMER} currently contains a module to compare FITS headers to the config file.
    \item \textbf{``Remembering'' the previous selection of \addressresponse{multi-source registration}.} This feature would limit manual user input by finding the two stars in the image as per the \addressresponse{multi-source} method described above, determining the relative positions of the two stars, and using the information from the first user click to determine which of the two stars is the primary in successive frames.
    \item \textbf{Reducing contrast error due to nearby stellar companions.} As mentioned in Section~\ref{sec:contrast curves}, nearby stellar companions induce contrast error. Introducing an option to mask out the companion's region (in separation and in azimuthal angle) in the contrast curve calculation  would provide a more accurate estimate of image sensitivity in the other regions of the image. 
\end{itemize}

Longer-term features that would require further development include:

\begin{itemize}
    \item \textbf{Extending functionality to other instruments.} The desire to incorporate a variety of instruments drove much of the code design. While incorporating certain instruments would likely require further abstraction of the \texttt{Instrument} class to encapsulate the idiosyncrasies of each instrument,
    each subsequent abstraction and extension cycle would make the process of adding further instruments iteratively easier. \addressresponse{Extension to higher-precision instruments (e.g., NIRC2 on the Keck II adaptive optics system) will likely require more robust treatments of distortions and nonlinearities caused by both the imager and the adaptive optics system \citep[e.g.,][]{neichel2015deep}.}
    
    \item \textbf{Checking for saturated and \addressresponse{multi-source} cases instead of relying on user input.} A user may not realize that their image is saturated or that their binaries are too similar in magnitude, so having the pipeline check for these cases and alert the user would increase the overall robustness of the pipeline.
    
    \item \textbf{Improving the saturation registration mode.} A number of methods are available to speed up \texttt{Python} codes: parallelization, \texttt{Numba} \citep{10.1145/2833157.2833162}, and \texttt{Cython} \citep{behnel2011cython}, to name a few. If the speed of this mode is increased to the point where it is comparable to the \addressresponse{quick-look} method, then it will become the primary method by which the pipeline performs reductions. Additionally, future versions of this software will allow users to input the desired number of rotations per target.
\end{itemize}

\section{Summary}\label{sec:summary}
We have presented the first public release of \texttt{SImMER}, an open-source reduction code for point sources. In addition to standard practices of dark-subtraction, flat-fielding, and bad pixel management, image registration can be run in modes suitable for saturated stars or wide binaries. Our pipeline can currently be run on data from the ShARCS camera on the Shane 3-m telescope and the PHARO camera on the Hale 5.1-m telescope. 

To the community, we provide open-source code on GitHub, code documentation, installation guides, and tutorials for reducing ShARCS and PHARO data on Read the Docs, and ease of installation on PyPI and \texttt{conda}; and from the community, we invite adaptations, issue reporting, and general involvement.
\begin{acknowledgments}

We acknowledge funding support from the Hellman Family Faculty Fund, the Sloan Foundation, and the David and Lucile Packard Foundation (grant 2019-69648). A.B. Savel acknowledges funding from the Heising-Simons Foundation.

We thank the members of Prof. Courtney Dressing's research group for providing helpful feedback on the \texttt{SImMER} user experience. In particular, we thank Steven Giacalone for assisting as we benchmarked different contrast curve implementations. We gratefully acknowledge the staff at Lick and Palomar observatories for their work in operating and maintaining their respective instruments. \addressresponse{Furthermore, we thank the anonymous reviewer for their careful reading of this manuscript that greatly improved its quality.}

This research made use of Photutils, an Astropy package for
detection and photometry of astronomical sources \citep{larry_bradley_2020_4044744}.

This research has made use of NASA’s Astrophysics Data System Bibliographic Services.

\end{acknowledgments}

\facilities{Hale (PHARO infrared camera), Shane (ShARCS infrared camera)}

\software{\texttt{astropy} (\citealt{price2018astropy}), \texttt{Cerberus} (\url{https://github.com/pyeve/cerberus}), \texttt{emcee} (\citealt{emcee}), \texttt{Matplotlib} (\citealt{hunter2007matplotlib}),  \texttt{NumPy} (\citealt{harris2020array}),
\texttt{pandas}  (\citealt{mckinney2010data}), \texttt{photutils} (\citealt{larry_bradley_2020_4044744}) \texttt{scikit-image} (\citealt{boulogne2014scikit}), \texttt{SciPy} (\citealt{virtanen2020scipy}), \texttt{tqdm} (\citealt{da2019tqdm})}%\clearpage \c

%% thebibliography produces citations in the text using \bibitem-\citealt
%% to the KEY in the \citealt commands (see the first section above).
%% place of the \citealt commands.



% \makeatletter
% \renewcommand\@biblabel[1]{}
% \makeatother
\bibliographystyle{aasjournal}
\bibliography{ref.bib}

\appendix
\section{Module descriptions}\label{sec:appendix}
We briefly describe each of the modules contained in \texttt{SImMER} below.

\begin{itemize}
    \item \texttt{add\_dark\_exp}: ShARCS observations often end the night with automated darks exposures. This module extracts header information from these darks and appends it to a log sheet.
    \item \texttt{analyze\_image}: Includes a compact analysis pipeline (source-finding, FWHM estimation, contrast curve calculation) that can be run after data reduction.
    \item \texttt{check\_logsheet}: Ensures that the log sheet is formatted correctly for subsequent parsing.
    \item \texttt{contrast}: Computes contrast curves.
    \item \texttt{create\_config}: Creates a formatted config file from an observer-input log sheet.
    \item \texttt{darks}: Combines darks images.
    \item \texttt{flats}: Combines flats images.
    \item \texttt{image}: Creates stacks of images, wrapping the darks and flats functionality.
    \item \texttt{insts}: Includes instrument-specific data and functions.
    \item \texttt{make\_triceratops\_contrasts}: Gathers contrast curves and formats them to be used alongside the \texttt{triceratops} \citep{giacalone2020vetting} code.
    \item \texttt{plotting}: Controls and produces plots.
    \item \texttt{registration}: Performs all image registration.
    \item \texttt{run\_night}: Wrapper to run all of \texttt{SImMER} for a single night of observations.
    \item \texttt{scipy\_utils}: Includes deprecated \texttt{SciPy} functionality.
    \item \texttt{search\_headers}: Searches the headers of data files for incomplete or missing information.
    \item \texttt{sky}: Includes functionality for skies.
    \item \texttt{summarize}: Wrapper to produce summary plots for a night of data reduction.
    \item \texttt{utils}: Utility functions for reading and slicing files.
    \item \texttt{validate\_config}: Confirms that the values listed in a configuration file match the values in the FITS file headers.
\end{itemize}



\end{document}